\documentclass[journal]{IEEEtran}
\usepackage{cite}
\usepackage{soul} 
\usepackage{url}
\usepackage{amsmath,amssymb,amsfonts}
\usepackage{array}
\usepackage{blkarray} 
\usepackage{xcolor}
\usepackage{graphicx}
\usepackage[caption=false,font=footnotesize]{subfig}

\begin{document}

\title{Error-Mitigated Multi-Layer Quantum Routing}


\author{
\IEEEauthorblockN{Wenbo Shi\IEEEauthorrefmark{1}\thanks{Corresponding author: Wenbo Shi (email: wenbo.shi@unsw.edu.au).}, Neel Kanth Kundu\IEEEauthorrefmark{2}\IEEEauthorrefmark{3}, and Robert Malaney\IEEEauthorrefmark{1}}\\
\IEEEauthorblockA{\IEEEauthorrefmark{1}School of Electrical Engineering and Telecommunications,
\textit{University of New South Wales}, Sydney, NSW, Australia.}
\IEEEauthorblockA{\IEEEauthorrefmark{2}Centre for Applied Research in Electronics, \textit{Indian Institute of Technology Delhi}, New Delhi, India.}
\IEEEauthorblockA{\IEEEauthorrefmark{3}Department of Electrical and Electronic Engineering, \textit{University of Melbourne, Melbourne}, Victoria, Australia.}}


\maketitle

\begin{abstract}

Due to the numerous limitations of current quantum devices, quantum error mitigation methods become potential solutions for realizing practical quantum applications in the near term. Zero-Noise Extrapolation (ZNE) and Clifford Data Regression (CDR) are two promising quantum error mitigation methods. Based on the characteristics of these two methods, we propose a new method named extrapolated CDR (eCDR). To benchmark our method, we embed eCDR into a quantum application, specifically multi-layer quantum routing. Quantum routers direct a quantum signal from one input path to a quantum superposition of multiple output paths  and are considered important elements of future quantum networks. Multi-layer quantum routers extend the scalability of quantum networks by allowing for further superposition of paths. We benchmark the performance of multi-layer quantum routers implemented on current superconducting quantum devices instantiated with the ZNE, CDR,  and eCDR methods. Our experimental results show that the new eCDR method significantly outperforms ZNE and CDR for the 2-layer quantum router. Our work highlights how new mitigation methods built from different components of pre-existing methods, and designed with a core application in mind, can lead to significant performance enhancements.

\end{abstract}


\section{Introduction}

Current quantum devices are commonly referred to as Noisy Intermediate-Scale Quantum (NISQ) devices due to their high intrinsic error rates, short coherence time, and limited   physical connections between qubits~\cite{brooks2019beyond, bharti2022noisy} -
 issues that lead to quantum error correction being non-implementable  on them~\cite{Brandhofer2021NISQ, Qin_2022}.
As an alternative, quantum error mitigation, a method that requires classical post-processing and additional executions of ancillary quantum circuits, is proposed as a potential pathway to near-term quantum advantages on NISQ devices~\cite{kandala2019error, takagi2022fundamental, Cai2023QEM}.

One  promising quantum error mitigation method is Zero-Noise Extrapolation (ZNE)~\cite{digitalZNE2020, zne2020He, Turk2022learningZNE, Krebsbach2022Optimization}, which extrapolates zero-noise results from noisy results obtained from ancillary quantum circuits.
These ancillary circuits are constructed by artificially introducing noise to a quantum circuit of interest (denoted as the original circuit), with each ancillary circuit possessing a distinct noise level.
Another promising mitigation method is Probabilistic Error Cancellation (PEC)~\cite{PEC2017shortdepth, pec2021Mari, van2023probabilistic}, in which  noiseless quantum gates are represented as linear combinations of noisy implementable quantum gates.
Yet another promising method is Clifford Data Regression (CDR)~\cite{Czarnik2021errormitigation, vnCDR2021, Strikis2021LearningBased, Perez2024extensionCDR}, which involves executing a group of near-Clifford circuits both on a noiseless simulator and a quantum device, generating a linear regression model based on the noiseless and noisy results.
The experimental results  collected from the original circuit run on the quantum device are then mitigated using the linear regression model.


Not withstanding the success of the above mitigation techniques shown in many studies, it is evident that in some applications their performance, at best, is limited; in some circumstances leading to worse outcomes relative to unmitigated circuits~\cite{Bultrini2023unifying, vnCDR2021, pec2021Mari, Majumdar2023Bestzne, Pascuzzi2022Computationally}. 
Therefore, room for new mitigation techniques, tailored to specific applications, exists. 
In this work, we propose a new quantum error mitigation method, named extrapolated CDR (eCDR), that builds a conceptual bridge between the ZNE and CDR  - exploiting the characteristics of both methods best suited to the application at hand. 
Our approach highlights how new error mitigation techniques - designed  around a specific application - can lead to significant performance improvement relative to ``off-the-shelf'' mitigation techniques.

The application we focus on to design and test our new mitigation method is one related to quantum routing~\cite{behera2019designing, Christensen2020coherentRouter}; 
A quantum router places an input data signal (classical or quantum) into a superposition of multiple output paths~\cite{yuan2015experimental, bartkiewicz2018implementation}; an outcome which can enhance the transmission of   data~\cite{kristj2023quantumnetworkscoherentrouting, Enhanced2018Ebler_quantumswitch} as well as form a basis for quantum random access memory~\cite{Arunachalam2015QRAM,Matteo2020QRAM, Xu2023QRAM}.
The quantum routing application can be benchmarked in terms of fidelity when embedded into Quantum State Tomography (QST), a technique used to determine unknown quantum states.
Our previous work~\cite{globecom2023wenbo} shows that applying ZNE and PEC in a concatenated manner (one method wholly embedded within the other) improves the entanglement fidelity of the quantum router, thereby demonstrating the feasibility of quantum routing on current quantum devices.
However, this concatenated method yields inferior results for more complex circuits such as those required for multi-layer quantum routers (in which each router output path is input into another router).
A new error mitigation method is required for this multi-layer space - a solution for which we provide here.  As we shall see our new method, eCDR, goes beyond a simple concatenation of different existing methods.




The remainder of this paper is as follows.
In Section~\ref{QEM}, we present the working principles of the ZNE, CDR, and eCDR methods.
In Section~\ref{QR2}, we introduce the circuits of multi-layer quantum routers and discuss the QST that we embedded into them for testing purposes. 
In this same section, using a 127-qubit NISQ device provided by IBM, we also report on experimental results of the quantum error mitigation methods applied to multi-layer quantum routers, 
comparing the performance with and without these methods.
Section~\ref{Conclusions} concludes our work.

\section{Quantum Error Mitigation} \label{QEM}

Before introducing the quantum error mitigation methods, we first clarify some necessary variables and notations.
Assume an original quantum circuit of interest $U$ includes $n'$ qubits and $K$ unitary gates $\{G_k\}_{k=1}^{K}$, which specify the qubit or qubits they are applied to. 
The original circuit, $U$, is given by 
$U = G_K \cdots G_k \cdots G_2 G_1$.
In this work, except where specifically defined, the measurement results of an observable $O$ applied to $U$ refer to the $Z$-basis measurement outcomes of $n \leq n'$ qubits in $U$.
Specifically, the ``measurement results'' represent the probabilities of all possible measurement outcomes occurring in a total of $C$ measurements.
We further denote the ``experimental results'' as the measurement results obtained from the quantum device.
Suppose the output state of $U$ is denoted by a density matrix $\sigma$, the expectation value of $O$ given $\sigma$ is 
$\left< O \right> = \text{Tr}[\sigma O]$.
Note that 
$\left< O \right>$ equals the sum of the probabilities of each measurement outcome multiplied by its corresponding eigenvalue.

\subsection{Overview of ZNE}

ZNE involves two main steps: constructing noise-scaled circuits and extrapolating estimated values to a zero-noise level.
To generate the noise-scaled circuits, some of the unitary gates in $U$ are randomly selected and then noise-scaled following
\begin{equation}
G_k \rightarrow G_k \left(G_k^\dag G_k\right)^{\xi}, \quad 
\xi = 0, 1, 2, \cdots
\text{.}
\end{equation}
This folding method artificially inserts noise without changing the effect of $U$ since $G_k^\dag G_k = I$, where $I$ is the identity operation.
Note that this noise-scaling method, known as gate folding, is not unique to the generation of noise-scaled circuits.
A similar method, known as circuit folding, applies the same folding logic but to the entire unitary circuit: $U \rightarrow U \left(U^\dag U\right)^{\xi}$.
If control pulses for realizing $G_k$ are accessible, the pulse stretching method, which extends noise by controlling the pulse duration for each unitary gate, could also be considered for generating noise-scaled circuits~\cite{digitalZNE2020}.
It can be found that the critical assumption of ZNE is that the noise in the quantum device can be described by noise-scale factors.
This implies that incoherent errors are the dominant type of noise and other types of errors are negligible, since only incoherent errors can be effectively amplified by these noise-scaling methods.

The noise-scale factors, denoted as 
$\{\lambda_j \}_{j=1}^{J}$, 
are utilized to quantify the levels of noise present in the noise-scaled circuits.
Specifically, $\lambda_j$ is defined as the ratio of the quantity of the unitary gates in the noise-scaled circuit to $K$, 
where $\lambda_j \geq 1$ and typically, $\lambda_1 = 1$.
To proceed with ZNE, the noise-scaled circuits, denoted as $\{ U_{\lambda_j}\}_{j=1}^{J}$, are executed on the quantum device for collecting their experimental results.
Note that $U_{\lambda_1} = U$, when $\lambda_1 = 1$, 
and the experimental results are used to calculate noisy expectation values $\tilde{\left< O \right>}_{\lambda_j}^{zne}$.
Based on the noisy expectation values, construction of functions relating the results to the noise levels is undertaken (e.g via the least-squares method).
Various extrapolation models, including polynomial extrapolation, can be considered in creating these functions (parameter-fitting to specific models).
The extrapolation models are functions of $\Lambda$, where $\Lambda \in \{ \lambda_0, \lambda_1, \cdots, \lambda_J\}$ and $\lambda_0 = 0$.
By evaluating the function at $\Lambda = \lambda_0$, an expectation value at zero-noise level, denoted as $\hat{\left< O \right>}_{\lambda_0}^{zne}$, is extrapolated.
Note that the extrapolation model can also be applied to the experimental results directly to obtain the zero-noise-level measurement results.



\subsection{Overview of CDR}


The main idea of CDR involves constructing near-Clifford circuits, denoted as $\{V_m\}_{m=1}^{M}$, that can be efficiently computed by classical simulators.
Note that these near-Clifford circuits are close to $U$ and evaluated both on a noiseless simulator and the quantum device.
The measurement results of $\{V_m\}$ obtained from the simulator are considered noiseless results, while those from the quantum device are considered noisy results.
Using these noiseless and noisy results, the noiseless and noisy expectation values of $O$ after executing the near-Clifford circuits are calculated, and a linear regression model is constructed in the form of 
\begin{equation} \label{eq:poly1}
\hat{\left< O \right>}^{cdr} = 
a \tilde{\left< O \right>} + b
\text{,}
\end{equation}
where $\hat{\left< O \right>}^{cdr}$ represents the CDR error-mitigated expectation value and $\tilde{\left< O \right>}$ represents the experimental expectation value of 
$O$ after executing $U$.
Note that $a$ and $b$ are real parameters determined by
\begin{equation}
(a, b) = 
\operatorname*{arg\,min}_{(a, b)} \sum_{m=1}^{M} 
\left[ \left< O \right>_m^{cdr} 
- \left(a \tilde{\left< O \right>}_m^{cdr}  + b\right) \right]^2
\text{,}
\end{equation}
where $\left\{\left< O \right>_m^{cdr}\right\}$ and $\left\{\tilde{\left< O \right>}_m^{cdr}\right\}$ represent the noiseless and noisy expectation values of $O$ after executing the near-Clifford circuits, respectively. 
The linear regression model can also be constructed directly from the experimental results of the near-Clifford circuits and then applied to the experimental results of $U$. Note that CDR can, in principle, perfectly mitigate the noise of a global depolarizing channel~\cite{Czarnik2021errormitigation}.


\begin{figure*}[t]
    \centering
    \includegraphics[width =\linewidth]{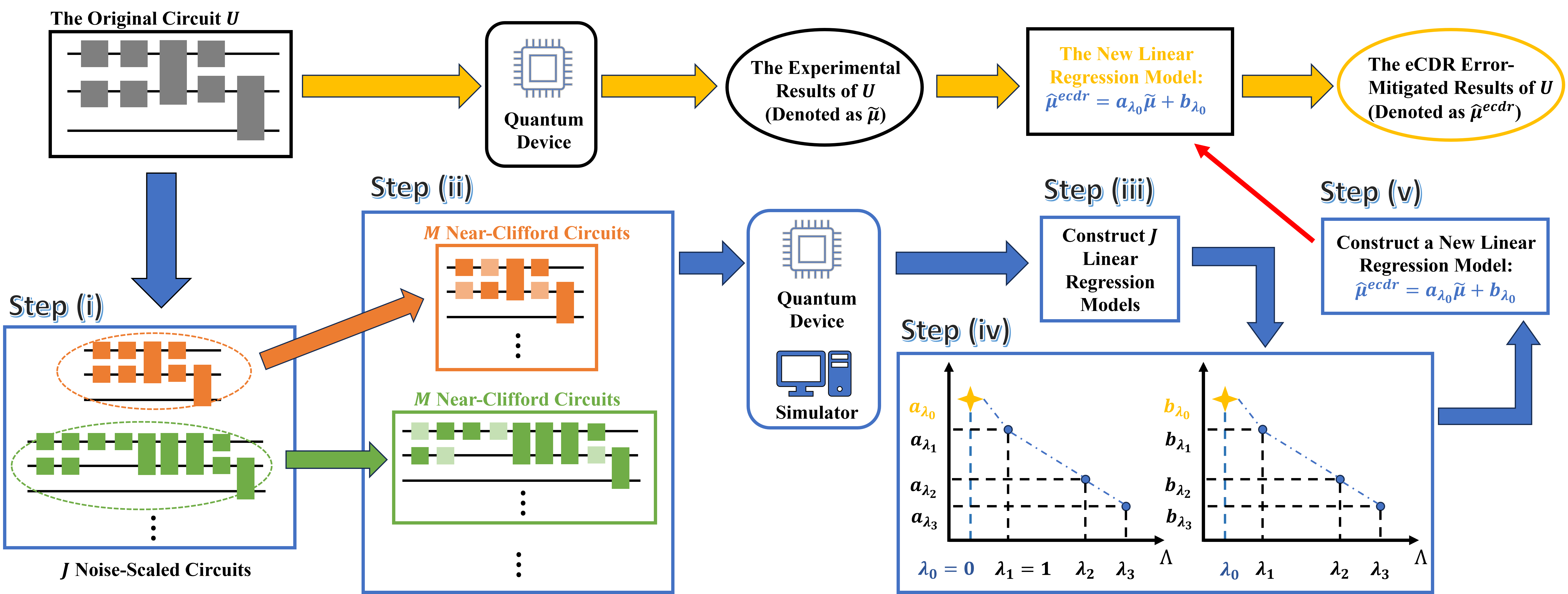}
\caption{Schematic of the eCDR method.
Based on $U$, $J$ noise-scaled circuits are generated, and $M$ near-Clifford circuits are constructed for each noise-scaled circuit. 
All near-Clifford circuits are executed on a simulator and a quantum device to construct a total of $J$ linear regression models.
Then, the real parameters in these models are extrapolated to build a new linear regression model for mitigating the experimental results of $U$.
}
    \label{fig:eCDRstr}
\end{figure*}

The concatenation of ZNE and CDR  denoted as ``ZNE+CDR'' and ``CDR+ZNE,'' can be considered for potentially better results.
The ZNE+CDR method first generates a group of noise-scaled circuits and then utilizes CDR to mitigate the experimental results of each noise-scaled circuit. 
These mitigated results are then utilized to extrapolate the zero-noise result.
The drawback of ZNE+CDR is that mitigating the experimental results of the noise-scaled circuits makes it challenging to redefine the relationship between these results and the noise scale factors of the noise-scaled circuits.
The main idea of CDR+ZNE is similar but in the opposite order: first, near-Clifford circuits close to $U$ are generated, and the experimental results of each near-Clifford circuit are error-mitigated by ZNE.
Afterward, this ZNE error-mitigated data is utilized to conduct the remaining steps of CDR: constructing a linear regression model and using this model for error mitigation.
The disadvantage of CDR+ZNE is that since this method finally relies on the relationship between noisy and noiseless results to mitigate errors, applying any error mitigation to the near-Clifford data disrupts this relationship.

\subsection{The eCDR Method}

\begin{figure*}[t]
    \centering
    \includegraphics[width =\textwidth]{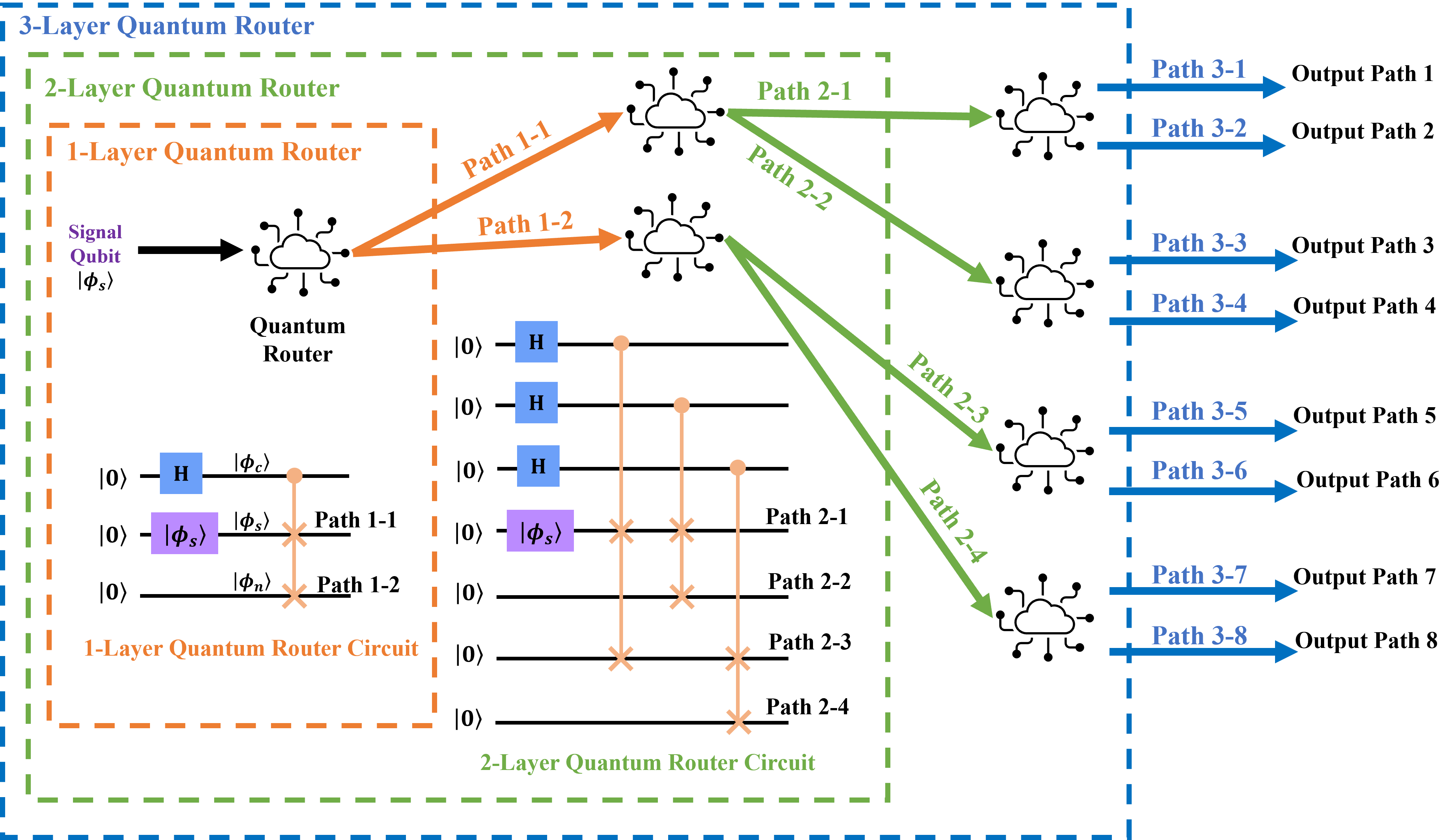}
\caption{Schematic of multi-layer quantum routers with 1- and 2-layer quantum router circuit.
In the circuits, all qubits are initialized in the $\vert 0 \rangle$ state.
The single-qubit gate in blue represents the Hadamard gate, while the one in purple prepares the signal qubit by transforming $\vert 0 \rangle$ into $\vert \phi_s\rangle$.
The 3-qubit gate is a controlled-swap gate that swaps the states of the qubits marked with cross symbols when the qubit marked with a solid circle is in the $\vert 1\rangle$ state and leaves them unchanged when it is in the $\vert 0\rangle$ state.
}
    \label{fig:QRstr}
\end{figure*}

The key idea of eCDR is that it establishes several groups of near-Clifford circuits. 
Each group is evaluated at different noise levels, and their experimental results are used to construct linear regression models. 
The real parameters in these models are then extrapolated to derive a new linear regression model, which is then applied to the experimental results of the original quantum circuit.
Note that eCDR can be used to derive error-mitigated expectation values,
however, here we use the experimental results to provide an example of the eCDR method.
The schematic of eCDR is illustrated in Fig.~\ref{fig:eCDRstr}, and its main steps are as follows.

(i)~\textit{Generation of noise-scaled circuits.} 
Firstly, we generate $J$ noise-scaled circuits, denoted as $\{ N_{\lambda_j}\}_{j=1}^{J}$, in the same manner as the initial step of ZNE.

(ii)~\textit{Generation of near-Clifford circuits.} 
Then, for each noise-scaled circuit, $N_{\lambda_j}$, we generate $M$ near-Clifford circuits, denoted as $\{T_{\lambda_j, m} \}_{m=1}^{M}$.
Note that the $M$ circuits in $\{T_{\lambda_j, m} \}$ are all close to $N_{\lambda_j}$ but slightly different from each other, as they are generated by randomly convert a portion of non-Clifford gates in $N_{\lambda_j}$ to Clifford gates.
There are a total of $J$ sets of $\{T_{\lambda_j, m} \}$ and a total of $D = JM$ near-Clifford circuits.

(iii)~\textit{Construction of linear regression models.} 
We execute $D$ near-Clifford circuits on both the simulator and the quantum device.
We denote the measurement results of $\{T_{\lambda_j, m} \}$ from the simulator and the quantum device as $\{ \mu_{\lambda_j, m}^{ecdr}\}$ and $\{ \tilde{\mu}_{\lambda_j, m}^{ecdr}\}$, respectively.
Note that $\{ \mu_{\lambda_j, m}^{ecdr}\}$ and $\{ \tilde{\mu}_{\lambda_j, m}^{ecdr}\}$ are regarded as noiseless and noisy measurement results, respectively.
We then construct a linear regression model in the form of
\begin{equation}
\hat{\mu}_{\lambda_j, m}^{ecdr} = 
a_{\lambda_j} \tilde{\mu}_{\lambda_j, m}^{ecdr} + b_{\lambda_j} 
\text{,}
\end{equation}
where $\hat{\mu}_{\lambda_j, m}^{ecdr}$ is an estimated measurement result of $T_{\lambda_j, m}$.
Note that $a_{\lambda_j}$ and $b_{\lambda_j}$ are real parameters determined by the least-squares method given by
\begin{equation}
(a_{\lambda_j}, b_{\lambda_j}) = 
\operatorname*{arg\,min}_{(a_{\lambda_j}, b_{\lambda_j})} 
\sum_{m=1}^{M} 
\left[\mu_{\lambda_j, m}^{ecdr} - (a_{\lambda_j} \tilde{\mu}_{\lambda_j, m}^{ecdr} + b_{\lambda_j} ) \right]^2
\text{.}
\end{equation}
There are in total of $J$ linear regression models corresponding to $J$ noise-scaled circuits.
This step is proposed based on CDR, which utilizes a linear regression model to describe the relationship between noiseless and noisy measurement results.

(iv)~\textit{Extrapolations.} 
We then collect all of the real parameters in these $J$ linear regression models, specifically $\{ a_{\lambda_j}\}$ 
and $\{ b_{\lambda_j}\}$.
We then employ curve fitting to these parameters to extrapolate two new parameters, denoted as $a_{\lambda_0}$ and $b_{\lambda_0}$.
Similar to ZNE, alternative extrapolation methods can be considered, we choose to use quadratic polynomial extrapolation model in our experiments.
Using $\{ a_{\lambda_j}\}$ as an example, the quadratic polynomial extrapolation is in the form of 
\begin{equation} \label{eq:poly2}
f(\Lambda) = c_0 + c_1 \Lambda + c_2 \Lambda^2
\text{,}
\end{equation}
where 
$c_0$, $c_1$, and $c_2$ are real parameters selected by 
\begin{equation}
(c_0, c_1, c_2) = 
\operatorname*{arg\,min}_{(c_0, c_1, c_2)} \sum_{j=1}^{J} 
\left[ a_{\lambda_j} - \left(c_0 + c_1 \lambda_j + c_2 \lambda_j ^2 \right) \right]^2
\text{.}
\end{equation}
This curve fitting process for $\{ b_{\lambda_j}\}$ is the same but the selected parameters, $c_0$, $c_1$, and $c_2$, are different.
By setting $\Lambda = \lambda_0$ in Eq.~\eqref{eq:poly2}, we determine $a_{\lambda_0}$ and $b_{\lambda_0}$,
which can also be expressed as
\begin{equation}
a_{\lambda_0}  = 
\sum_{j = 1}^{J} \gamma_j a_{\lambda_j} \quad
\text{and} \quad
b_{\lambda_0}  = 
\sum_{j = 1}^{J} \gamma_j b_{\lambda_j}
\text{,}
\end{equation}
where $\{\gamma_j\}$ are real parameters determined by the values of $\{\lambda_j\}$ and the selected extrapolation model.

(v)~\textit{Construction of a new linear regression model for mitigation.} 
With $a_{\lambda_0}$ and $b_{\lambda_0}$, we construct a new linear regression model given by 
\begin{equation} \label{eq:ecdr}
\hat{\mu}^{ecdr} = a_{\lambda_0} \tilde{\mu} + b_{\lambda_0}
\text{,}
\end{equation}
where $\tilde{\mu}$ is the experimental results of $U$ and $\hat{\mu}^{ecdr}$ is the corresponding eCDR error-mitigated results.
Similar to CDR, the selection of this linear regression model in eCDR is motivated by considering the effect of the global depolarizing channel (see Appendix).

It is worth mentioning that the eCDR method should not be confused with variable-noise CDR (vnCDR)~\cite{vnCDR2021, Bultrini2023unifying}.
Although both vnCDR and eCDR use noise-scaled near-Clifford data for error mitigation, the approaches to utilizing this data differ.
In vnCDR, a set of near-Clifford circuits, denoted as $\{\mathcal{T}_{m}\}_{m=1}^M$, are first generated, and then, for each $\mathcal{T}_{m}$, a set of noise-scaled near-Clifford circuits, denoted as $\{\mathcal{T}_{m, \lambda_j} \}_{j=1}^{J}$, are generated.
After the execution of $D$ noise-scaled near-Clifford circuits, vnCDR constructs an extrapolation model for mitigation given by
\begin{equation} \label{eq:vncdr}
\hat{\mu}^{vncdr} = 
\sum_{j=1}^{J} A_j \cdot \tilde{\mu}_{\lambda_j}
\text{,}
\end{equation}
where $\hat{\mu}^{vncdr}$ stands for the vnCDR error-mitigated result of $U$ and $\tilde{\mu}_{\lambda_j}$ represents the experimental result of $U_{\lambda_j}$.
Note that $\tilde{\mu}_{\lambda_1} = \tilde{\mu}$ with $\lambda_1 = 1$ and $A_j$ are parameters selected by the least-squares method following
\begin{equation}
A_j = 
\operatorname*{arg\,min}_{A_j} \sum_{m=1}^{M} 
\left[ \mu_m^{vncdr} - \sum_{j=1}^{J} A_j \cdot \tilde{\mu}_{m, \lambda_j}^{vncdr} \right]^2
\text{,}
\end{equation}
where $\mu_{m}^{vncdr}$ represents the measurement results of $\mathcal{T}_{m}$ collected from the simulator and  $\tilde{\mu}_{m, \lambda_j}^{vncdr}$ represents the measurement results of $\mathcal{T}_{m, \lambda_j}$ collected from the quantum device.
The vnCDR and eCDR methods utilize the noise-scaled near-Clifford data to construct the extrapolation model and the new linear regression model for mitigation, respectively. 
By comparing Eqs.~\eqref{eq:ecdr} and~\eqref{eq:vncdr}, it can be observed that eCDR requires fewer resources compared to vnCDR: 
eCDR only requires the execution of $U$, while vnCDR requires the execution of $\{ U_{\lambda_j}\}$, which includes $J$ circuits.

For multi-layer quantum routers, the complexity of their circuits increases with the number of layers (see details in the next Section).
Specifically, the number of qubits and circuit depth of a router circuit grow with the number of layers of the quantum router, with more qubits becoming entangled in higher-layer quantum routers.
To mitigate errors in the router circuit with increased complexity, we designed eCDR, which combines the advantages of ZNE and CDR.
ZNE is mainly effective for incoherent errors, while CDR is primarily effective for coherent errors and measurement errors.
By extrapolating the parameters within the linear regression model, eCDR demonstrates the potential to effectively mitigate errors in the router circuits with increased complexity.

\section{Error-mitigated Quantum Routers} \label{QR2}
\subsection{Multi-layer Quantum Routers}


The simplest structure of the quantum router consists of a signal qubit $\vert \phi_s\rangle$, a control qubit $\vert \phi_c\rangle$, and an ancillary qubit, denoted as $\vert \phi_n\rangle = \vert0\rangle_n$.
The control qubit directs the signal qubit, which contains quantum information, to the desired output path.
We denote the signal and ancillary qubits as the path qubits.
Specifically, the signal qubit is given by  
$\vert \phi_s\rangle = \alpha_s \vert 0 \rangle_s + \beta_s \vert 1\rangle_s$
and the control qubit is given by 
$\vert \phi_c\rangle = \alpha_c \vert 0 \rangle_c + \beta_c \vert 1\rangle_c$,
where $\alpha_s$, $\beta_s$, $\alpha_c$, and $\beta_c$ are parameters satisfying $|\alpha_s|^2 + |\beta_s|^2 = 1$ and $|\alpha_c|^2 + |\beta_c|^2 = 1$.
When the control qubit is in the state $\vert 0\rangle_c$, the signal qubit is routed to the first output path, and when the control qubit is in the state $\vert 1\rangle$, the signal qubit is routed to the second output path.
When the control qubit is in a superposition, the output of the quantum router becomes an entanglement between the control qubit and the two output paths.
The output of the quantum router is in the form of
\begin{equation}
\vert\Phi\rangle_f =
\alpha_c \vert0\rangle_c \vert\phi_s\rangle_1 \vert\phi_n\rangle_2 
+ \beta_c \vert1\rangle_c \vert\phi_n\rangle_1 \vert\phi_s\rangle_2 
\text{,}
\end{equation}
where the subscripts $1$ and $2$ represent the first and second output paths, respectively.

We denote the quantum router with two output paths as the 1-layer quantum router.
To increase the number of output paths, we concatenate quantum routers as depicted in Fig.~\ref{fig:QRstr}.
The output paths of the first-layer (second-layer) quantum router serve as the input paths for the quantum routers in the second (third) layer.
The 2-layer quantum router consists of three 1-layer quantum routers, resulting in four output paths, and the 3-layer quantum router consists of seven 1-layer quantum routers, resulting in eight output paths.
For clarification, we denote the first output path of the 1-layer quantum router as path 1-1, and the remaining output paths following a similar notation.

The 1-layer quantum router circuit is also demonstrated in Fig.~\ref{fig:QRstr}.
In this router circuit, the signal qubit is prepared in a random quantum state using the purple single-qubit gate.
In our experiments, this purple gate transforms the signal qubit to a quantum state with the parameters $\alpha_s = 0.5+0.13i$ and $\beta_s = -0.82-0.22i$, where $i$ is the imaginary unit.
The control qubit is converted to a superposition with $\alpha_c = \beta_c = 1/\sqrt{2}$ using the Hadamard gate.
The controlled-swap gate, the 3-qubit gate in orange, realizes the quantum routing process.
The 2-layer router circuit is similar to the 1-layer router circuit but with 3 control qubits, 4 path qubits (1 signal qubit and 3 ancillary qubits), and 3 controlled-swap gates, as demonstrated in Fig.\ref{fig:QRstr}.
Similarly, the 3-layer router circuit has 7 control qubits, 8 path qubits (1 signal qubit and 7 ancillary qubits), and 7 controlled-swap gates.

\subsection{Multi-layer Quantum Routers with QST}

We choose signal fidelity $F$ as our performance metric, defined by
\begin{equation}
F = \left( \text{Tr} \sqrt{\sqrt{\rho} \rho' \sqrt{\rho}}   \right)^2
\text{,}
\end{equation}
where $\rho = \vert \phi_s \rangle \langle \phi_s\vert$ and $\rho'$ is the reconstructed quantum state of the signal qubit at the output of the quantum router.
In other words, $\rho'$ represents the noisy experimental density matrix of the signal qubit, while $\rho$ is the noiseless density matrix used for comparison.

\begin{figure}[t]
    \centering
    \includegraphics[width =\linewidth]{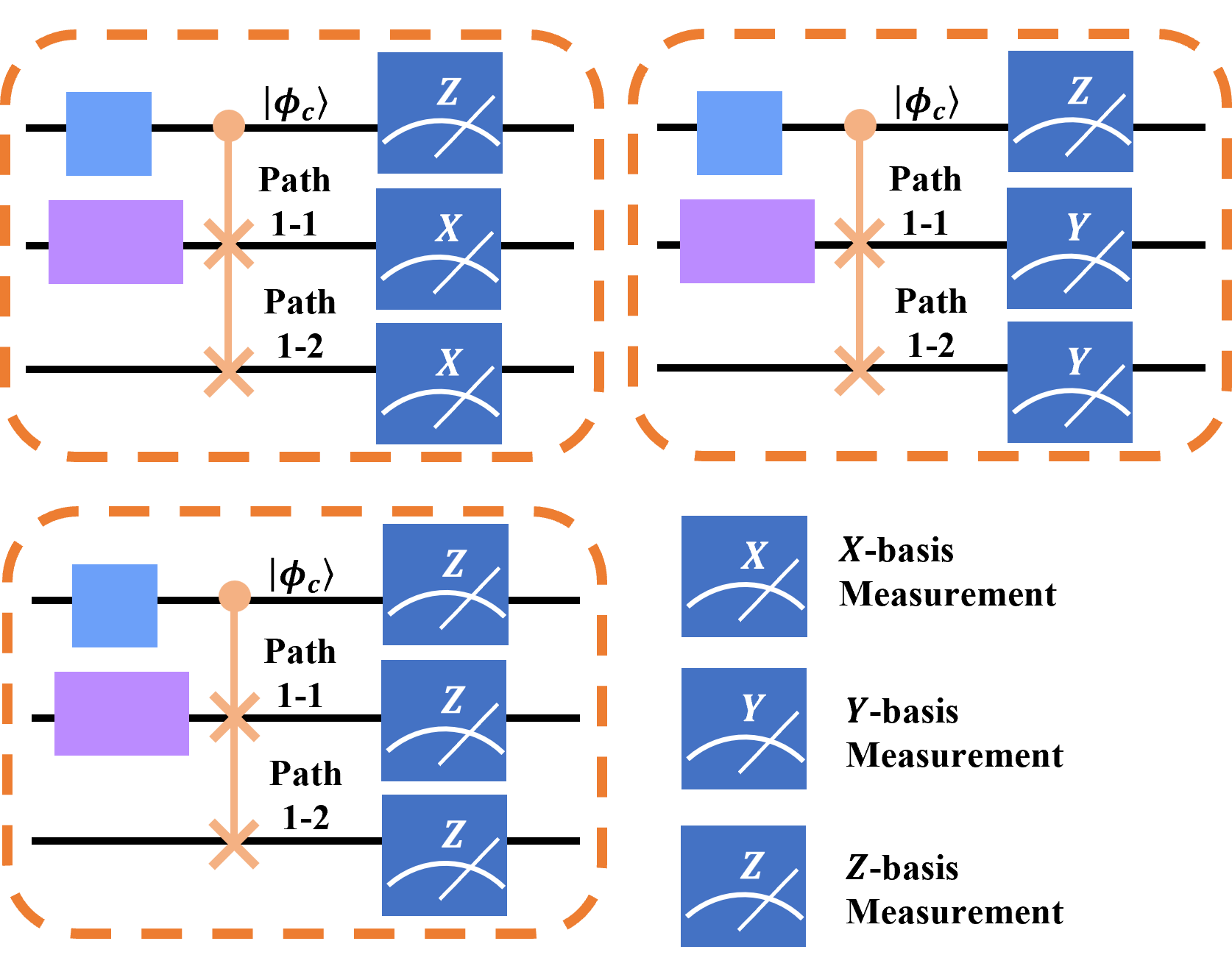}
\caption{Tomography circuits utilized to reconstruct the quantum state of the signal qubit in the 1-layer quantum router.
}
    \label{fig:QSTstr}
\end{figure}

QST is utilized for the reconstruction of $\rho'$, which is a 1-qubit density matrix given by
\begin{equation} \label{eq:1q_qst}
\rho' = 
\frac{1}{2} \bigl( S_0 I + S_1 X + S_2 Y + S_3 Z \bigr)
\text{,}
\end{equation}
where 
$I = \begin{pmatrix}
1 & 0\\
0 & 1
\end{pmatrix}$, 
$X= \begin{pmatrix}
0 & 1\\
1 & 0
\end{pmatrix}$, 
$Y = \begin{pmatrix}
0 & -i\\
i & 0
\end{pmatrix}$, and 
$Z = \begin{pmatrix}
1 & 0\\
0 & -1
\end{pmatrix}$.
Note that $S_0$, $S_1$, $S_2$, and $S_3$ are parameters determined by the results of $X$\mbox{-}, $Y$\mbox{-}, and $Z$-basis measurements.
Specifically, these parameters can be expressed as
\begin{equation} \label{eq:Pr}
\begin{split}
S_0 &= P_{Z0} + P_{Z1} = 1 \text{,} \quad
S_1 = P_{X0} -P_{X1} \text{,} \\
S_2 &= P_{Y0} - P_{Y1} \text{,} \quad \text{and } \,
S_3 = P_{Z0} - P_{Z1}
\text{,}
\end{split}
\end{equation} 
where $P_{Z0}$ and $P_{Z1}$ represent the probabilities of obtaining the states $\vert0\rangle$ (the $+1$ eigenvalue) and $\vert1\rangle$ (the $-1$ eigenvalue) in the $Z$-basis measurement.
Similarly, $P_{X0}$, $P_{X1}$, $P_{Y0}$, and $P_{Y1}$ represent the corresponding probabilities in the $X$- and $Y$-basis measurements~\cite{altepeter2005photonic}.
With these measurement results of the signal qubit, we reconstruct $\rho'$ using Eq.~\eqref{eq:1q_qst} and~\eqref{eq:Pr}.


Since the signal information can be found in multiple output paths after the quantum routing process, we apply the $Z$-basis measurements to the control qubits, whose measurement result indicates the location of the signal qubit.
We also apply the three basis measurements to each path qubit and post-select only the measurement results of the path qubit that contains the signal information to reconstruct $\rho'$.
For the 1-layer quantum router, there are a total of three tomography circuits, as demonstrated in Fig.~\ref{fig:QSTstr}, which correspond to the 1-layer router circuit with three different measurement operators: $O = Z \otimes Z \otimes Z$, $Z \otimes X \otimes X$, and $Z \otimes Y \otimes Y$.
Since the quantum device only supports the $Z$-basis measurements, we perform the $X$-basis measurements by adding a Hadamard gate before the $Z$-basis measurement and achieve the $Y$-basis measurements by sequentially adding an $S^{\dagger}$ gate (which induces a $-\pi/2$ phase) and a Hadamard gate before the $Z$-basis measurement.

\subsection{Experimental Setup of eCDR}

In this work, we utilize a 127-qubit device named \textit{ibm\_sherbrooke}~\cite{ibmq} and a simulator to conduct our experiments.
The \textit{ibm\_sherbrooke} is currently one of the smallest quantum devices provided by IBM.
The simulator is realized through IBM's open-source software development kit---the Quantum Information Science toolKit (Qiskit)~\cite{qiskit2024}.
Our experiments utilize 3, 7, and 15 qubits to implement 1\mbox{-}, 2\mbox{-}, and 3-layer quantum routers, respectively.
The tomography circuits must undergo transpilation, a process that converts them into transpiled circuits, prior to their execution on the quantum device.
Note that the transpilation is also realized through Qiskit.
In transpilation, the control and path qubits are mapped to specific physical qubits of the quantum device, and each quantum gate in the tomography circuits is decomposed into basis gates, which can be directly implemented on the quantum devices.
In addition, Mitiq software package~\cite{LaRose2022mitiqsoftware} is partially utilized for the implementation of ZNE and CDR.

In our experiments, $U$ corresponds to the unitary parts of the transpiled circuit, \textit{i.e.}, the transpiled circuit excluding measurements.
Any quantum circuit that requires execution on the quantum device or the simulator is executed $C = 20,000$ times.
The ZNE, CDR, and vnCDR methods were originally proposed to be utilized for expectation values, while in our experiments, we implement modified versions of them for calculating the signal fidelity.
Specifically, we apply these three methods to the measurement results instead of the expectation values of $O$  (henceforth,
the terms ZNE, CDR, and vnCDR  will refer  only  measurement result usage).

We now discuss experimental setups of the six mitigation methods we investigate.
(i)~For ZNE, we generate three noise-scaled circuits (with measurements reintroduced) with $\lambda_j$ values of approximately 1, 3, and 5, respectively.
We choose the quadratic polynomial extrapolation model (as shown in Eq.~\eqref{eq:poly2}) to extrapolate error-mitigated values.
(ii)~For CDR, we generate $50$ near-Clifford circuits that approximate the transpiled circuit to obtain adequately near-Clifford data for error mitigation.
The transpiled circuit only contains the basis gates and measurement operations, and among all of the basis gates, only the $Rz(\varphi)$ gate could potentially be a non-Clifford gate, depending on the value of $\varphi$. 
Note that the $Rz(\varphi)$ gate rotates a single-qubit along the $Z$-axis, where $\varphi$ is a phase factor.
We randomly select non-Clifford gates in the transpiled circuit with a probability of $90\%$ to ensure that the near-Clifford circuits can be executed efficiently on the simulator, even for more complex transpiled circuits. 
The selected non-Clifford gates are converted to the nearest Clifford gates by adjusting the value of $\varphi$.
(iii)~For eCDR, we generate the same noise-scaled circuits as in ZNE, and for each noise-scaled circuit, we generate ten near-Clifford circuits in the perspective of resource-saving.
(iv)~For ZNE+CDR, we first generate three noise-scaled circuits, as in ZNE, for each transpiled circuit. We then generate five near-Clifford circuits, as in CDR, for each noise-scaled circuit.
There are a total of 45 quantum circuits to be executed, as there are three transpiled circuits.
The CDR error-mitigated results become the new results for the noise-scaled circuits and are then further mitigated by the quadratic polynomial extrapolation model.
(v)~For CDR+ZNE, we use only the first transpiled circuit (the one with the measurement operator $O = Z^{\otimes 3}$) to generate five near-Clifford circuits. Based on each of these, we generate three noise-scaled circuits.
There are a total of 15 quantum circuits to be executed.
(vi)~Finally, for vnCDR, we use the same approach as eCDR to generate ten near-Clifford circuits, and for each near-Clifford circuit we generate three noise-scaled circuits with $\lambda_j$ values of approximately 1, 3, and 5.
The generation of the error mitigation model in vnCDR follows its original version: the model is generated based on the expectation values of $\{\mathcal{T}_{m, \lambda_j}\}$.
However, the model is then applied to the measurement results of $\{U_{\lambda_j}\}$ to obtain error-mitigated fidelity.

Due to the complexity of the quantum routers with a higher number of layers, we simplify the 3-layer router circuit to enhance the accuracy of the measurement results.
We simplify the 3-layer router circuit by randomly selecting four control qubits to be in the superposition while setting the remaining control qubits to be in the state $\vert0\rangle_c$.
Based on this setup for the control qubits, only certain path qubits are expected to contain the signal information at the output of the quantum router (these will be measured in the three basis).
Among the control qubits, only those in superposition will be measured in the $Z$-basis. 

\subsection{Experimental Results of eCDR}

\begin{figure}[t]
    \centering
    \includegraphics[width =\linewidth]{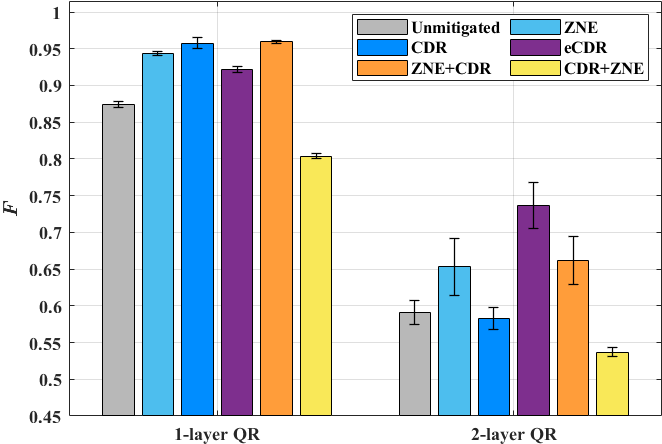}
\caption{Signal fidelity, $F$, of the 1- and 2-layer quantum router with and without quantum error mitigation methods realized on the \textit{ibm\_sherbrooke} device. 
QR stands for quantum router.
Note that the fidelity results for Unmitigated, ZNE, CDR, and eCDR are averaged over 20 repetitions, and the other fidelity results (ZNE+CDR and CDR+ZNE) are averaged over 3 repetitions.
The error bars represent the standard deviation.
}
    \label{fig:Fbar}
\end{figure}

The experimental results of the 1- and 2-layer quantum routers with ZNE, CDR, eCDR, ZNE+CDR, and CDR+ZNE methods are demonstrated in Fig.~\ref{fig:Fbar}.
The unmitigated experimental results are also illustrated for comparison.
Each bar shows the averaged $F$ from 20 repetitions (Unmitigated, ZNE, CDR, and eCDR) or 3 repetitions (ZNE+CDR and CDR+ZNE), and the error bars represent the standard deviation. 
Although these error mitigation methods demonstrate similar error-mitigating performance for the 1-layer quantum routers, eCDR yields significantly enhanced results compared to the other four methods for the 2-layer quantum router.

For the 3-layer quantum router, the unmitigated experimental result in terms of fidelity is approximately $0.5$, and the ZNE, CDR, and eCDR methods are basically ineffective.
These 3-layer mitigation results indicate that mitigating errors for quantum circuits with large circuit depths remains a significant challenge.
One potential direction in meeting this challenge could be a combination of quantum error mitigation with quantum error correction techniques. However, any introduction of quantum error correction will almost certainly require an advancement in current hardware to lower noise levels.

In addition to the five quantum error mitigation methods mentioned above, we also evaluated the performance of the ZNE+PEC method (the concatenation of ZNE and PEC) and the vnCDR method when applied to the quantum routers.
The ZNE+PEC method slightly improves $F$ for the 1-layer quantum router compared to the unmitigated fidelity, while it reduces $F$ for the 2-layer quantum router. 
The vnCDR error-mitigated signal fidelities 
(averaged from 2 repetitions) 
for the multi-layer quantum routers are all slightly lower than the corresponding unmitigated signal fidelities. 
We found that the mitigation model (as described by Eq.~\eqref{eq:vncdr}) generated in vnCDR varies significantly across repetitions.

The vnCDR method generates the mitigation model using noisy expectation values, and this model is then applied to measurement results for mitigation.
The mitigation model in vnCDR is formed from near-Clifford data and can also be identified as an extrapolation model, which is typically employed in ZNE and constructed from noise-scaled experimental data.
These mismatches could be the main reasons for the poor performance of vnCDR.

For ZNE, additional errors are more prone to be introduced to the noise-scaled circuits, $\{U_{\lambda_j}\}$, with higher complexity and larger value of $\lambda_j$ when executed on the quantum device. 
As a result, $\lambda_j$ may not accurately represent the noise ratio between the measurement results of $U_{\lambda_j}$ and those of $U$, thereby decreasing the effectiveness of error mitigation for the 2-layer quantum router.
Our eCDR method employs an extrapolated linear regression model (as shown in Eq.~\eqref{eq:ecdr}) to estimate the relationship between noisy and noiseless measurement results of $U$.
If the linear regression model generated by CDR can precisely describe the relationship — usually occurring when unmitigated measurement results of $U$ are close to the noiseless measurement results — further introducing extrapolations (which is the main step of eCDR) can add additional errors, resulting in worse outcomes.
However, due to the fact that the unmitigated measurement results of the 2-layer router circuit are relatively noisy, the extrapolated linear regression model in eCDR provides a more accurate description of the relationship, rather than introducing additional errors.
In summary, ZNE and CDR are more effective for simpler quantum circuits with relatively accurate unmitigated measurement results of $U$, while eCDR is more suitable for mitigating errors in quantum circuits with higher complexity and relatively noisy unmitigated measurement results.


\section{Conclusions} \label{Conclusions}

In this work, we proposed a quantum error mitigation method, denoted as eCDR, which conceptually combines the characteristics of two promising error mitigation methods, ZNE and CDR.
We embedded ZNE, CDR, and eCDR methods into the 1\mbox{-}, 2\mbox{-}, and 3-layer quantum routers to benchmark their performance conducted on a 127-qubit quantum device named \textit{ibm\_sherbrooke}.
For the 1-layer quantum router, the three methods demonstrate a similar positive mitigation effect, whereas for the 2-layer quantum router, the eCDR method demonstrates superior performance compared to the other two methods.
For the 3-layer quantum router, error mitigation was found to be ineffective.
Our results indicate, in the context of quantum routing, the circuit depths above which  error mitigation will be successful using current hardware.


\section*{Acknowledgment}
We acknowledge the use of IBM~Quantum services for this work and the advanced services provided by IBM~Quantum Hub at the University of Melbourne. 
WS is supported by the China Scholarship Council, the University of New South Wales, and the Sydney Quantum Academy, Sydney, NSW, Australia.
NKK acknowledges the support from the INSPIRE Faculty Fellowship awarded by the Department of Science and Technology, Government of India (Reg. No.: IFA22‐ENG 344) and the New Faculty Seed Grant from the Indian Institute of Technology Delhi.

\appendix[\normalfont Motivation for Linear Functions in eCDR]\label{Appdx:eCDRmath}

Consider a global depolarizing channel is applied to $\sigma$, the output state of $U$, followed by a measurement of the observable $O$.
This depolarizing channel $\mathcal{E}$ is given by
\begin{equation}
\mathcal{E} (\sigma) = 
\left(1- \epsilon \right) \sigma + \epsilon I / d
\text{,}
\end{equation}
where $d = 2^{n'}$ is the Hilbert-space dimension and $\epsilon$ is a parameter that describes the noise, ranging from 0 to 1. 
In terms of expectation value, the effect of the depolarizing channel leads to 
\begin{equation} \label{eq:channeleffect}
\text{Tr}\left[\mathcal{E}(\sigma) O \right] = 
\left(1- \epsilon \right) \text{Tr}\left[ \sigma O \right] 
+ \epsilon \frac{\text{Tr}[O]}{d}
\text{.}
\end{equation}
The noisy expectation value of $O$ after executing $U$ is $ \tilde{\left< O \right>} = \text{Tr}\left[\mathcal{E}(\sigma) O \right]$ and the corresponding noiseless expectation value is $ \left< O \right> = \text{Tr}\left[ \sigma O \right]$, leading to
\begin{equation} 
\tilde{\left< O \right>} = 
\left(1- \epsilon \right) \left< O \right>
+ \epsilon \frac{\text{Tr}[O]}{d}
\text{.}
\end{equation}
The error-mitigated result of eCDR in terms of expectation value is given by
\begin{equation} \label{eq:ecdrFull}
\begin{split}
\hat{\left< O \right>}^{ecdr} &= 
a_{\lambda_0} \tilde{\left< O \right>} + b_{\lambda_0} \\
&= 
\sum_{j = 1}^{J} \gamma_j a_{\lambda_j} \tilde{\left< O \right>} + \sum_{j = 1}^{J} \gamma_j b_{\lambda_j}
\text{,}
\end{split}
\end{equation}
where $\hat{\left< O \right>}^{ecdr}$ is the eCDR error-mitigated expectation value.
To completely mitigate the effect of the global depolarizing channel, \textit{i.e.}, to achieve $\hat{\left< O \right>}^{ecdr} = \left< O \right>$, eCDR needs to satisfy
\begin{equation} \label{eq:ecdrDep1}
\begin{split}
a_{\lambda_0} & = 
\sum_{j = 1}^{J} \gamma_j a_{\lambda_j}
= \frac{1}{1-\epsilon} \\
\text{and} \quad
b_{\lambda_0} &= 
\sum_{j = 1}^{J} \gamma_j b_{\lambda_j}
= - \frac{\epsilon \text{Tr}[O]}{(1-\epsilon) d}
\text{.}
\end{split}
\end{equation}

Assuming the global depolarizing channel is applied to the original circuit, $U$, $j$ times, the output state, $\sigma$, then becomes 
\begin{equation}
\mathcal{E}^{j} (\sigma) = 
\left(1- \epsilon \right)^j \sigma + \left[1- (1-\epsilon)^j \right] I / d
\text{.}
\end{equation}
In terms of expectation value, the above expression can also be expressed as 
\begin{equation}
\text{Tr}\left[\mathcal{E}^{j} (\sigma) O \right] 
= \left(1- \epsilon \right)^{j} \text{Tr}\left[ \sigma O \right] 
+ \left[1- (1-\epsilon)^j \right] \frac{\text{Tr}[O]}{d}
\text{,}
\end{equation}
which is equivalent to 
\begin{equation}
\tilde{\left< O \right>}  = 
\left(1- \epsilon \right)^{j} \left< O \right>
+ \left[1- (1-\epsilon)^j \right] \frac{\text{Tr}[O]}{d}
\text{.}
\end{equation}
Therefore, the eCDR error-mitigated expectation value is given by
\begin{equation}
\begin{split}
& \hat{\left< O \right>}^{ecdr} = \\
& \sum_{j = 1}^{J} \gamma_j 
\left[ a_{\lambda_j} \left(1- \epsilon \right)^{j} \left< O \right>
+ a_{\lambda_j}  \left[1- (1-\epsilon)^j \right] \frac{\text{Tr}[O]}{d} + b_{\lambda_j} \right] 
\text{.}
\end{split}
\end{equation}
We see that $ \hat{\left< O \right>}^{ecdr}$ is equivalent to $\left< O \right>$ when
\begin{equation} \label{eq:ecdrDepj}
\begin{split}
& \sum_{j = 1}^{J} \gamma_j a_{\lambda_j}  \left(1- \epsilon \right)^{j} = 1  \quad \text{and} \\
& \sum_{j = 1}^{J} \gamma_j \left[ a_{\lambda_j}  \left[1- (1-\epsilon)^j \right] \frac{\text{Tr}[O]}{d} + b_{\lambda_j} \right] 
 = 0
\text{.}
\end{split}
\end{equation}
Therefore, we can state that eCDR is capable of completely mitigating global depolarizing noise across distinct noise levels.
Note that the above analysis is provided based on Refs.~\cite{Czarnik2021errormitigation, vnCDR2021}.
However, our restrictions of the parameters, as shown in Eqs.~\eqref{eq:ecdrDep1} and~\eqref{eq:ecdrDepj}, are derived based on the eCDR method.



%


\bibliographystyle{IEEEtran}
\bibliography{IEEEabrv,References}

\begin{thebibliography}{10}
\providecommand{\url}[1]{#1}
\csname url@samestyle\endcsname
\providecommand{\newblock}{\relax}
\providecommand{\bibinfo}[2]{#2}
\providecommand{\BIBentrySTDinterwordspacing}{\spaceskip=0pt\relax}
\providecommand{\BIBentryALTinterwordstretchfactor}{4}
\providecommand{\BIBentryALTinterwordspacing}{\spaceskip=\fontdimen2\font plus
\BIBentryALTinterwordstretchfactor\fontdimen3\font minus \fontdimen4\font\relax}
\providecommand{\BIBforeignlanguage}[2]{{%
\expandafter\ifx\csname l@#1\endcsname\relax
\typeout{** WARNING: IEEEtran.bst: No hyphenation pattern has been}%
\typeout{** loaded for the language `#1'. Using the pattern for}%
\typeout{** the default language instead.}%
\else
\language=\csname l@#1\endcsname
\fi
#2}}
\providecommand{\BIBdecl}{\relax}
\BIBdecl

\bibitem{brooks2019beyond}
M.~Brooks, ``Beyond quantum supremacy: {The} hunt for useful quantum computers,'' \emph{Nature}, vol. 574, no. 7776, pp. 19--22, 2019.

\bibitem{bharti2022noisy}
K.~Bharti, A.~Cervera-Lierta, T.~H. Kyaw, T.~Haug, S.~Alperin-Lea, A.~Anand, M.~Degroote, H.~Heimonen, J.~S. Kottmann, T.~Menke, W.-K. Mok, S.~Sim, L.-C. Kwek, and A.~Aspuru-Guzik, ``Noisy intermediate-scale quantum algorithms,'' \emph{Reviews of Modern Physics}, vol. 94, 015004, 2022.

\bibitem{Brandhofer2021NISQ}
S.~Brandhofer, S.~Devitt, T.~Wellens, and I.~Polian, ``Special session: {Noisy Intermediate-Scale Quantum (NISQ)} computers—{How} they work, how they fail, how to test them?'' in \emph{2021 IEEE 39th VLSI Test Symposium (VTS)}, 2021, pp. 1--10.

\bibitem{Qin_2022}
D.~Qin, X.~Xu, and Y.~Li, ``An overview of quantum error mitigation formulas,'' \emph{Chinese Physics B}, vol.~31, no. 9, 090306, 2022.

\bibitem{kandala2019error}
A.~Kandala, K.~Temme, A.~D. C{\'o}rcoles, A.~Mezzacapo, J.~M. Chow, and J.~M. Gambetta, ``Error mitigation extends the computational reach of a noisy quantum processor,'' \emph{Nature}, vol. 567, no. 7749, pp. 491--495, 2019.

\bibitem{takagi2022fundamental}
R.~Takagi, S.~Endo, S.~Minagawa, and M.~Gu, ``Fundamental limits of quantum error mitigation,'' \emph{NPJ Quantum Information}, vol.~8, no. 1, 114, 2022.

\bibitem{Cai2023QEM}
Z.~Cai, R.~Babbush, S.~C. Benjamin, S.~Endo, W.~J. Huggins, Y.~Li, J.~R. McClean, and T.~E. O'Brien, ``Quantum error mitigation,'' \emph{Reviews of Modern Physics}, vol.~95, no. 4, 045005, 2023.

\bibitem{digitalZNE2020}
T.~Giurgica-Tiron, Y.~Hindy, R.~LaRose, A.~Mari, and W.~J. Zeng, ``Digital zero noise extrapolation for quantum error mitigation,'' in \emph{2020 IEEE International Conference on Quantum Computing and Engineering (QCE)}, 2020, pp. 306--316.

\bibitem{zne2020He}
A.~He, B.~Nachman, W.~A. de~Jong, and C.~W. Bauer, ``Zero-noise extrapolation for quantum-gate error mitigation with identity insertions,'' \emph{Physical Review A}, vol. 102, no. 1, 012426, 2020.

\bibitem{Turk2022learningZNE}
P.~Turk and A.~Ozaki, ``Learning zero noise extrapolation for deterministic quantum circuits,'' in \emph{2022 IEEE International Conference on Quantum Computing and Engineering (QCE)}, 2022, pp. 265--274.

\bibitem{Krebsbach2022Optimization}
M.~Krebsbach, B.~Trauzettel, and A.~Calzona, ``Optimization of {R}ichardson extrapolation for quantum error mitigation,'' \emph{Physical Review A}, vol. 106, no. 6, 062436, 2022.

\bibitem{PEC2017shortdepth}
K.~Temme, S.~Bravyi, and J.~M. Gambetta, ``Error mitigation for short-depth quantum circuits,'' \emph{Physical Review Letters}, vol. 119, no. 18, 180509, 2017.

\bibitem{pec2021Mari}
A.~Mari, N.~Shammah, and W.~J. Zeng, ``Extending quantum probabilistic error cancellation by noise scaling,'' \emph{Physical Review A}, vol. 104, no. 5, 052607, 2021.

\bibitem{van2023probabilistic}
E.~Van Den~Berg, Z.~K. Minev, A.~Kandala, and K.~Temme, ``Probabilistic error cancellation with sparse {P}auli - {L}indblad models on noisy quantum processors,'' \emph{Nature Physics}, vol.~19, no.~8, pp. 1116--1121, 2023.

\bibitem{Czarnik2021errormitigation}
P.~Czarnik, A.~Arrasmith, P.~J. Coles, and L.~Cincio, ``Error mitigation with {C}lifford quantum-circuit data,'' \emph{{Quantum}}, vol. 5, 592, 2021.

\bibitem{vnCDR2021}
A.~Lowe, M.~H. Gordon, P.~Czarnik, A.~Arrasmith, P.~J. Coles, and L.~Cincio, ``Unified approach to data-driven quantum error mitigation,'' \emph{Physical Review Research}, vol.~3, no. 3, 033098, 2021.

\bibitem{Strikis2021LearningBased}
A.~Strikis, D.~Qin, Y.~Chen, S.~C. Benjamin, and Y.~Li, ``Learning-based quantum error mitigation,'' \emph{PRX Quantum}, vol.~2, no. 4, 040330, 2021.

\bibitem{Perez2024extensionCDR}
J.~Pérez-Guijarro, A.~Pagès-Zamora, and J.~R. Fonollosa, ``Extension of {C}lifford data regression methods for quantum error mitigation,'' in \emph{2024 IEEE International Conference on Acoustics, Speech and Signal Processing (ICASSP)}, 2024, pp. 9691--9695.

\bibitem{Bultrini2023unifying}
D.~Bultrini, M.~H. Gordon, P.~Czarnik, A.~Arrasmith, M.~Cerezo, P.~J. Coles, and L.~Cincio, ``Unifying and benchmarking state-of-the-art quantum error mitigation techniques,'' \emph{{Quantum}}, vol. 7, 1034, 2023.

\bibitem{Majumdar2023Bestzne}
R.~Majumdar, P.~Rivero, F.~Metz, A.~Hasan, and D.~S. Wang, ``Best practices for quantum error mitigation with digital zero-noise extrapolation,'' in \emph{2023 IEEE International Conference on Quantum Computing and Engineering (QCE)}, vol.~1, 2023, pp. 881--887.

\bibitem{Pascuzzi2022Computationally}
V.~R. Pascuzzi, A.~He, C.~W. Bauer, W.~A. de~Jong, and B.~Nachman, ``Computationally efficient zero-noise extrapolation for quantum-gate-error mitigation,'' \emph{Physical Review A}, vol. 105, no. 4, 042406, 2022.

\bibitem{behera2019designing}
B.~K. Behera, T.~Reza, A.~Gupta, and P.~K. Panigrahi, ``Designing quantum router in {IBM} quantum computer,'' \emph{Quantum Information Processing}, vol.~18, pp. 1--13, 2019.

\bibitem{Christensen2020coherentRouter}
K.~S. Christensen, S.~E. Rasmussen, D.~Petrosyan, and N.~T. Zinner, ``Coherent router for quantum networks with superconducting qubits,'' \emph{Physical Review Research}, vol.~2, no. 1, 013004, 2020.

\bibitem{yuan2015experimental}
X.~Yuan, J.-J. Ma, P.-Y. Hou, X.-Y. Chang, C.~Zu, and L.-M. Duan, ``Experimental demonstration of a quantum router,'' \emph{Scientific Reports}, vol.~5, no. 1, 12452, 2015.

\bibitem{bartkiewicz2018implementation}
K.~Bartkiewicz, A.~{\v{C}}ernoch, and K.~Lemr, ``Implementation of an efficient linear-optical quantum router,'' \emph{Scientific Reports}, vol.~8, no. 1, 13480, 2018.

\bibitem{kristj2023quantumnetworkscoherentrouting}
H.~Kristj{\'a}nsson, Y.~Zhong, A.~Munson, and G.~Chiribella, ``Quantum networks with coherent routing of information through multiple nodes,'' \emph{arXiv:2208.00480}, 2022.

\bibitem{Enhanced2018Ebler_quantumswitch}
D.~Ebler, S.~Salek, and G.~Chiribella, ``Enhanced communication with the assistance of indefinite causal order,'' \emph{Physical Review Letters}, vol. 120, no. 12, 120502, 2018.

\bibitem{Arunachalam2015QRAM}
S.~Arunachalam, V.~Gheorghiu, T.~Jochym-O’Connor, M.~Mosca, and P.~V. Srinivasan, ``On the robustness of bucket brigade quantum {RAM},'' \emph{New Journal of Physics}, vol.~17, no. 12, 123010, 2015.

\bibitem{Matteo2020QRAM}
O.~D. Matteo, V.~Gheorghiu, and M.~Mosca, ``Fault-tolerant resource estimation of quantum random-access memories,'' \emph{IEEE Transactions on Quantum Engineering}, vol.~1, pp. 1--13, 2020.

\bibitem{Xu2023QRAM}
S.~Xu, C.~T. Hann, B.~Foxman, S.~M. Girvin, and Y.~Ding, ``Systems architecture for quantum random access memory,'' in \emph{Proceedings of the 56th Annual IEEE/ACM International Symposium on Microarchitecture}, 2023, p. 526–538.

\bibitem{globecom2023wenbo}
W.~Shi and R.~Malaney, ``Error-mitigated quantum routing on noisy devices,'' in \emph{2023 IEEE Global Communications Conference}, 2023, pp. 5475--5480.

\bibitem{altepeter2005photonic}
J.~B. Altepeter, E.~R. Jeffrey, and P.~G. Kwiat, ``Photonic state tomography,'' \emph{Advances In Atomic, Molecular, and Optical Physics}, vol.~52, pp. 105--159, 2005.

\bibitem{ibmq}
\BIBentryALTinterwordspacing
``{IBM Q}uantum,'' 2021. [Online]. Available: \url{https://quantum.ibm.com/}
\BIBentrySTDinterwordspacing

\bibitem{qiskit2024}
A.~Javadi-Abhari, M.~Treinish, K.~Krsulich, C.~J. Wood, J.~Lishman, J.~Gacon, S.~Martiel, P.~D. Nation, L.~S. Bishop, A.~W. Cross, B.~R. Johnson, and J.~M. Gambetta, ``Quantum computing with {Q}iskit,'' \emph{arXiv:2405.08810}, 2024.

\bibitem{LaRose2022mitiqsoftware}
R.~LaRose, A.~Mari, S.~Kaiser, P.~J. Karalekas, A.~A. Alves, P.~Czarnik, M.~El~Mandouh, M.~H. Gordon, Y.~Hindy, A.~Robertson, P.~Thakre, M.~Wahl, D.~Samuel, R.~Mistri, M.~Tremblay, N.~Gardner, N.~T. Stemen, N.~Shammah, and W.~J. Zeng, ``Mitiq: {A} software package for error mitigation on noisy quantum computers,'' \emph{{Quantum}}, vol. 6, 774, 2022.

\end{thebibliography}

\end{document}